\documentclass[12pt]{article}

\usepackage{graphics}
\usepackage{graphicx}
\usepackage{epsfig}

\usepackage[toc,page]{appendix}

\usepackage{amsmath}
\usepackage{amssymb}
\usepackage{amsthm}
\usepackage{lscape}
\usepackage{multicol}
\usepackage[font=small,labelfont=bf]{caption}
\DeclareCaptionFont{tiny}{\tiny}

\def\b{\beta}

\newcommand{\be}{\begin{equation}}
\newcommand{\ee}{\end{equation}}
\newcommand{\bea}{\begin{eqnarray}}
\newcommand{\eea}{\end{eqnarray}}

\begin{document}

\hskip 12cm

\begin{center}
\center{\huge Quantum phase transition of high dimensional Yang-Mills theories}
\vskip 3cm
{\large N. Irges, G. Koutsoumbas and K. Ntrekis}
\vskip .5cm
 {\it Department of Physics, National Technical University of Athens\\
         GR-15780 Athens, Greece}

\vskip 2cm

{\bf Abstract}
\end{center}

We determine the critical value of the coupling where the first order quantum phase transition takes place
for lattice $SU(2)$ Yang-Mills theories in dimensions higher than four.
Within a Mean-Field approach we derive an approximate law valid for any dimension $d$ and in the context of a Monte Carlo approach,
in addition to the already known $d=5$ case, we look at $d=6,7,8.$

\newpage

Even though high ($d>4$) dimensional Yang-Mills theories are perturbatively non-renormalizable, one can not
exclude the possibility that there exists a regime in their phase diagram where a physically useful cut-off effective theory
can be constructed. Also from the theoretical point of view, these theories contain the basic ingredients (the gauge fields) of
various field theories motivated by string theory.\footnote{For Monte Carlo studies of high dimensional (supersymmetric) Yang-Mills theories
from the point of view of matrix models, see \cite{Kostas}.} Therefore, their study is potentially useful.

A possible universal property of high dimensional Yang-Mills theories that emerges from previous studies
comes from the fact that in five dimensions a "bulk" or "quantum" phase transition
appears dividing the confined phase from a Coulomb phase \cite{Creutz}. Recall that $4d$ pure Yang-Mills 
(for $SU(N)$ with $N<4$) \footnote{For $N\ge 4$ a first order bulk phase transition emerges already in $d=4$ \cite{SU4}.} at zero temperature
has only a confined phase where the gauge fields always form flux tubes.
Intuition says that as the number of dimensions transverse to the surface of the tube increases, the harder is to sustain a stable flux tube
thus requiring a stronger coupling. In phase diagram terminology, we expect the confined phase to persist but also to shrink as $d$ increases.
Here we will perform a first check of this statement. 

We take $SU(2)$ as our model and regularize it on a $d$-dimensional, Euclidean, periodic
lattice with lattice spacing $a$ and linear dimension $L$. The action is the standard Wilson plaquette action
\be
S_E^L =  \beta \sum_x \sum_{1 \le \mu,\nu \le d}\left[1-\frac{1}{2}
{\rm tr} \left\{ U_{\mu\nu}(x) \right\} \right],
\ee 
where $U_{\mu\nu}(x) = U_\mu(x)U_\nu(x+a
\hat{\mu})U_\mu^\dagger(x+a \hat{\nu})U_\nu^\dagger(x)$, $\mu,\nu=1,\cdots ,d$ is the elementary plaquette located at the site $x$
with $U_\mu(x) = e^{i a A_\mu(x)}$ and $\beta = \frac{4 a}{g^2}$
represents the dimensionless lattice coupling. 
We will employ two methods, on one hand the Mean-Field (MF) approximation \cite{DZ,MFtorus}, an analytic method expected to work well
near the phase transition and in general increasingly well as $d$ grows\footnote{Or when $N$ of $SU(N)$ grows; this however tends to shrink the Coulomb phase instead \cite{largeN}!}
and Monte Carlo (MC) simulations on the other.

For $SU(2)$ in $d$ dimensions, the (ungauge-fixed) MF approach to zeroth order determines the confined and Coulomb phases via the solution
to the coupled equations for the MF background $v_0$ \cite{DZ}
\be
{v_0}=\frac{I_2(h_0)}{I_1(h_0)}, \hskip 1cm h_0=2v_0^3 (d-1)\b\label{MFbck}
\ee
with $I_\nu(h_0)$ the modified Bessel function,
by defining the Coulomb phase as the regime of $\b$ where there is a solution with $v_0\ne 0$ and as the confined phase otherwise.
In \cite{MFtorus} the equations were solved for $d=5$ by a numerical, iterative method.
The smallest positive and real non-vanishing value of the background $v_{0c}$ satisfying eqs.(\ref{MFbck})
(i.e. where the iteration stabilizes), determines the critical value of the lattice coupling $\b_c$
where the phase transition takes place. It is expected to be a quantum phase transition since the MF
at this order is volume independent. This of course needs to be checked. In fact, for $d=5$ it was found by a MC simulation on a $4^5$ lattice in 1979 by Creutz
to be a quantum, first order phase transition. Subsequently this was confirmed (and extended to anisotropic lattices) by several authors \cite{5d}.
Apart from the fact that both methods agree on the order of the transition, their estimates for the value of the critical coupling are also quite close:
the $\b_c^{\rm MF} \simeq 1.6762017$ of the MF (corresponding to $v_{0c}\simeq 0.73333$) \cite{MFtorus} is to be compared with the $\b_c^{\rm MC} = 1.642\pm 0.015$ of the MC \cite{Creutz}.

An observation stemming from eqs.(\ref{MFbck}) is that the quantity
$B= (d-1)\b_c^{\rm MF}$ and therefore also $v_{0c}$ are
$d$-independent. We can then solve eqs.(\ref{MFbck}) for general $d$ by noting that
the zero of the function $F=I_2(h_0)/I_1(h_0)-v_0$ that signals the
phase transition is one where $F(v_0)$ has an extremum. 
The extremization condition $F'=0$ yields
\be
3Bv_0^2 \left[1-\frac{I_2}{I_1}\left(\frac{I_0}{I_1}+\frac{I_2}{I_1}\right)+\frac{I_3}{I_1}\right] = 1.\label{ext}
\ee
Using eq. (\ref{MFbck}) and the recursion identity $I_\nu(h_0) =  I_{\nu-2}(h_0) - \frac{2(\nu-1)}{h_0}
I_{\nu-1}(h_0)$ we can express $I_3/I_1=1-4v_0/h_0$ and $I_0/I_1=2/h_0+v_0$ and reduce 
eq. (\ref{ext}) to the quadratic equation
\be
x^2 - x + \frac{5}{3B} = 0
\ee
with $x=v_0^2$. 
The solution determines the value of the background at the critical point 
\be 
v_{0c} =
\frac{1}{\sqrt{2}} \sqrt{1\pm \sqrt{1-\frac{20}{3}\frac{1}{B}}}\,.\label{rel} 
\ee 
Substituting the above back in the equation $F=0$ results
in an algebraic expression with only parameter $B$, whose relevant
root can be found numerically to be $B\simeq 6.704840$, determining
$v_{0c}\simeq 0.7333$ from the upper sign of eq.(\ref{rel}), as
expected. Thus, we find that the equation \be (d-1)\b_c^{\rm
MF}\simeq 6.704840 \label{eqapp} \ee fixes the $SU(2)$ critical
coupling in any dimension $d>4$.

What makes it possible to go high in $d$ with Monte Carlo simulations is that we are dealing with a bulk phase transition.
This means that as long as the lattice extent is large enough so that finite size effects do not interfere,
the phase transition is visible. Most times a $4^d$ lattice will suffice to observe the effect, even though
larger lattices will be clearly needed to describe it with better precision.
The order parameter used in order to locate the phase transition is the plaquette
\be P = \frac{2}{d (d-1) L^D} \sum_x \sum_{1 \le \mu<\nu \le
d}\left[1-\frac{1}{2} {\rm tr} \left\{ U_{\mu\nu}(x) \right\} \right].\ee 
The Kennedy-Pendleton heat bath algorithm \cite{kenpen} has been used to update the gauge field. 
Far from the region of the phase transition the number of thermalization sweeps needed so that the plaquette achieves its equilibrium value has been 
of the order of 100, 150, 200 for $d=6,7,8$ respectively. We have also employed overrelaxation hits to decorrelate 
the measurements. They don't seem to affect the results very much, since the measurements of the autocorrelation 
times yield very small values.  The acceptance rates have been of the order of $70 - 90$ per cent.

The phase diagram has been obtained for $d=6,\ 7,\ 8,$ while we also quote the results for $d=5.$ The lattice sizes used
have a linear dimension $L=4$ and, after thermalization, hysteresis loops have been performed. The step in $\beta$ 
was $0.07$ (starting at $\beta = 0.40$ and going up to $\beta = 1.8$ and back); after 1000 initial heat up sweeps, 
$8000$ iterations through the lattice have been done at each $\beta$ value. 
There exist well known approximations that we have used as guides. In the strong coupling regime (small values of $\beta$) 
the plaquette $P$ is well approximated by $P \simeq \frac{\beta}{4},$ while in the weak coupling the approximations reads: 
$P \simeq 1-\frac{3}{d \beta}.$ It appears that the $\beta$ values shown in the figure are not large enough to approach 
the weak coupling limit, but we have checked that for larger values the agreement is good. The results for the hysteresis runs along 
with the strong and weak coupling approximations are shown on the left in Fig. 1. The phase transition is seen to be a strong 
first order one, even for this quite modest linear dimension of the lattices. We can read off estimates for the $\beta$ intervals 
in which the critical values lie: $1.29<\beta<1.45$ for $d=6,$ $1.07<\beta<1.27$ for $d=7,$ and $0.86<\beta<1.04$ for $d=8.$  
The right of Fig. 1 depicts a comparison of the mean field estimates for the critical $\beta$ and the corresponding analytical 
expression (\ref{eqapp}) against the results of the Monte Carlo runs. We have also quoted the five-dimensional result 
from \cite{Creutz} for completeness. We observe a quite good agreement between the two methods. 

%
\begin{figure}[!t]
\begin{minipage}{.49\textwidth}
\centerline{\epsfig{file=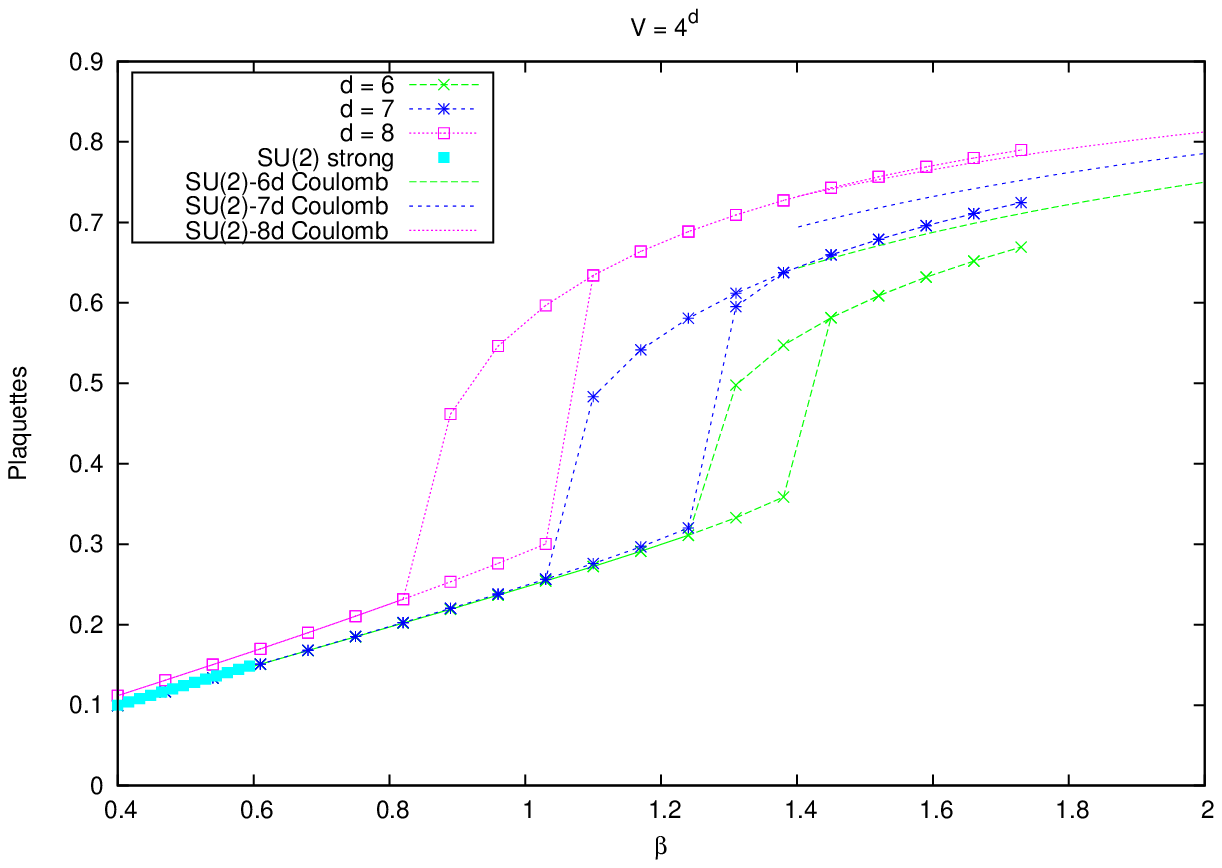,width=8.5cm}}
\end{minipage}
\begin{minipage}{.49\textwidth}
\centerline{\epsfig{file=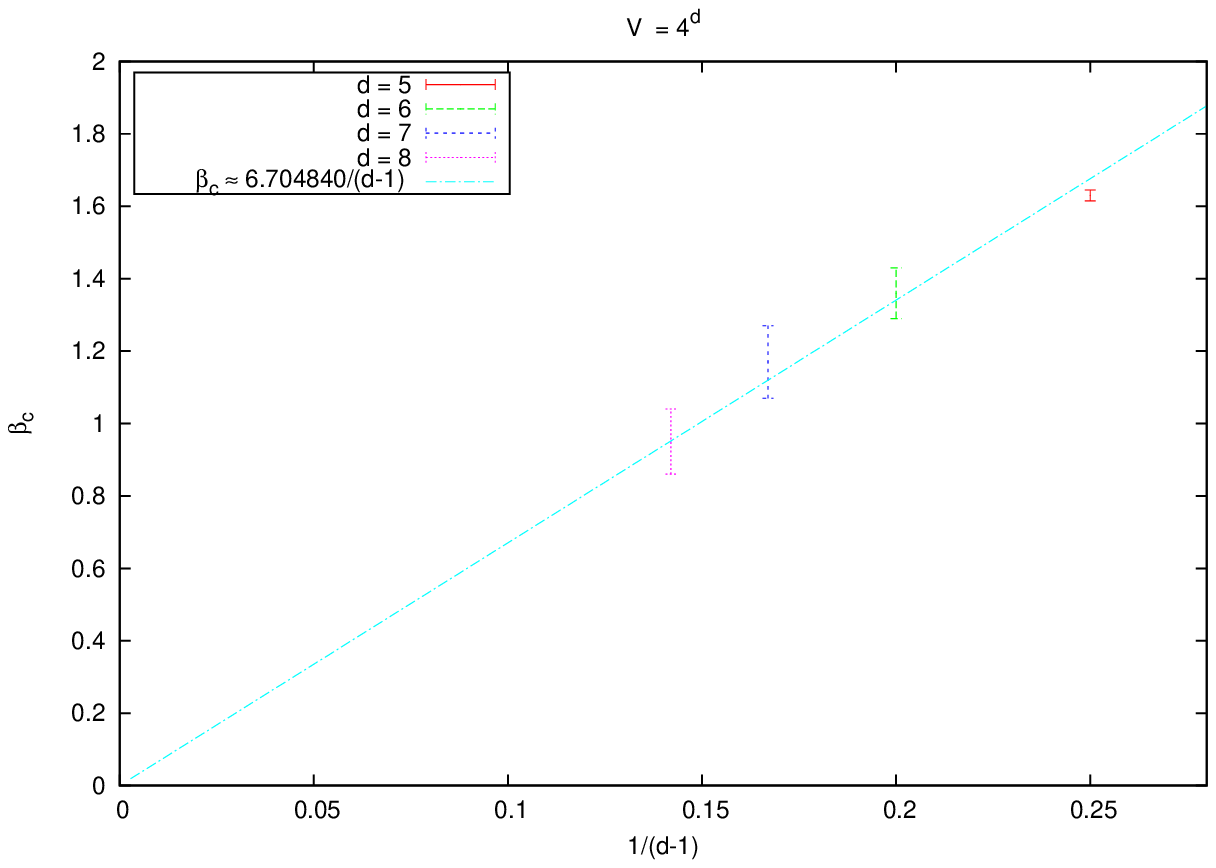,width=8.5cm}}
\end{minipage}
\caption{Left: Hysteresis loops for 6, 7 and 8 dimensions. 
The strong and weak coupling predictions are also included.
Right: The pseudocritical values for $\beta_c^{\rm MC}$ versus $\frac{1}{d-1}$ and their 
estimated errors. The Mean-Field prediction eq.(\ref{eqapp}) is represented by the straight line.
\label{fig}}
\end{figure}
%

The determination of the order of the phase transition as well as of the pseudocritical values for $\beta$ is usually done through 
long runs and by examining the fluctuations of the plaquette between the values pertaining to each of the two metastable states. 
This permits the construction of histograms and the determination of the volume dependence of the specific heats. Unfortunately, 
the gap between the two values grows so much with increasing dimensions, that it has proven impossible to observe this behavior 
with the algorithm used. This would require a multicanonical simulation, so we defer this to a forthcoming publication. 
For the time being, we have used long runs (of the order of $10^5$ iterations) just to be sure about the upper and lower bounds that we report. 
The results are encoded in the upper and lower bounds for the critical $\beta$ reported above.

Extending the work of \cite{Creutz} we determined the critical value of the coupling where 
a first order bulk phase transition takes place for high dimensional $SU(2)$ lattice gauge theories.
We first derived a law valid in any dimension $d>4$ based on the Mean-Field method and then performed corresponding Monte Carlo checks for the first time in $d=6,\  7$ and 8 dimensions.

The values of the plaquettes at selected values of $\beta,$ as well as several results relating to the error and autocorrelation issues are presented in the Appendix.

\vskip .1cm {\bf Acknowledgements} We would like to thank K. Farakos
for illuminating discussions and P. de Forcrand for useful comments. This research is implemented under the
ARISTEIA II action of the operational programme education and
long life learning and is co-funded by the European Union (European
Social Fund) and National Resources of Greece.


\section{Appendix}

We collect in this Appendix some details of the simulations. Near the region of the phase transition the autocorrelation question becomes important. For the six-dimensional models we plot in Figs. \ref{fig_6_129} the mean plaquette resulting in long runs for a hot and a cold start and values of $\beta$ smaller than the phase transition value. We see that the hot start makes the plaquettes take on their final values very soon, while the cold start spends considerable computer time in a false vacuum before actually landing on its true value. The time spent in the false vacuum becomes longer as one approaches the phase transition. For the relatively large $\beta's$ depicted in Figs.  \ref{fig_6_142} the opposite effect takes place: it is the hot start which gives a fluctuation around the false vacuum and this fact is more intense as one approaches the phase transition region; on the contrary, the cold start fluctuates around the true vacuum from the beginning. It appears that the measurements in this region are highly correlated; however the correlation functions indicate that the autocorrelation time is bigger than the total computer time of the run, so it makes no sense to display them. This behaviour just signals the limitations of the method. Similar results for $d=7$ are plotted in Figs. \ref{fig_7_107} and \ref{fig_7_131}. Similar behaviour is observed for $d=8,$ but one has to fine tune too much to achieve the corresponding wandering around the false vacua, so we don't show the results.

It is useful to report more details on the values of the plaquettes and their errors, as calculated by the Jackknife method. 
We report the relevant values at the values of $\beta$ that have been used for the determination of the phase diagram in tables 1, 2, 3 for $d=6,\ d=7,\ d=8$ respectively. We also depict the results of the long runs  in Fig. 6, including the error bars. We show both hot starts and cold starts along with the corresponding errors. The errors grow large only near the phase transition, as expected.

\begin{landscape}
\begin{figure}[t]
\begin{minipage}{.49\textwidth}
\centerline{\epsfig{file=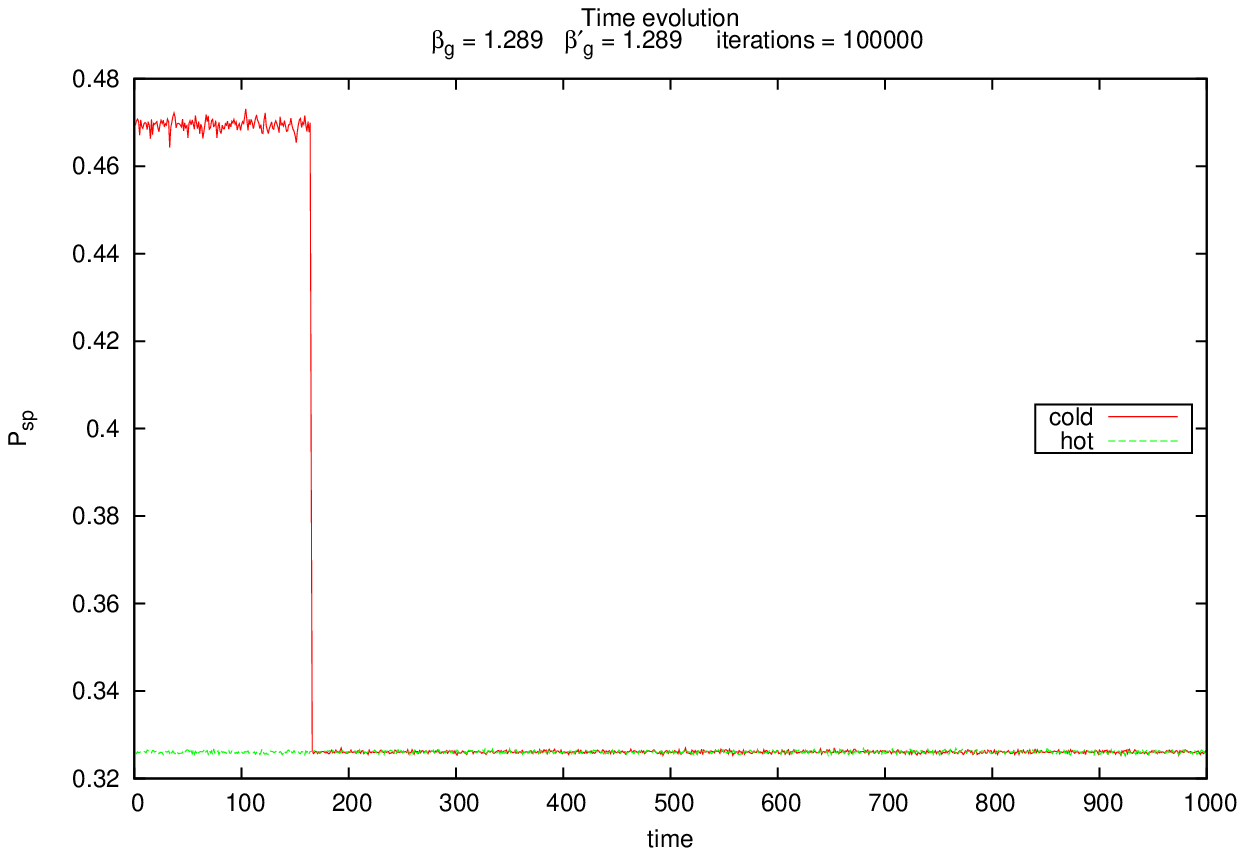,width=8.5cm}}
\end{minipage}
\begin{minipage}{.49\textwidth}
\centerline{\epsfig{file=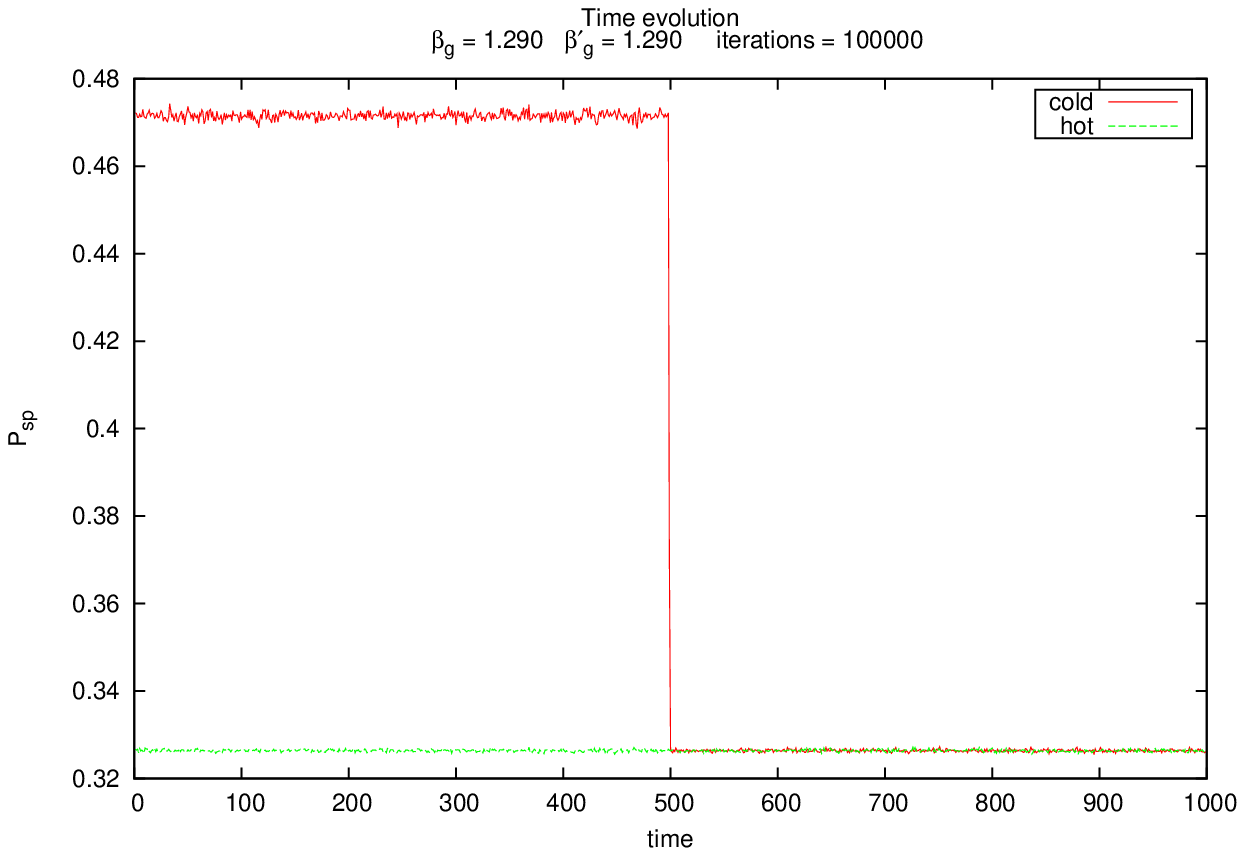,width=8.5cm}}
\end{minipage}
\begin{minipage}{.49\textwidth}
\centerline{\epsfig{file=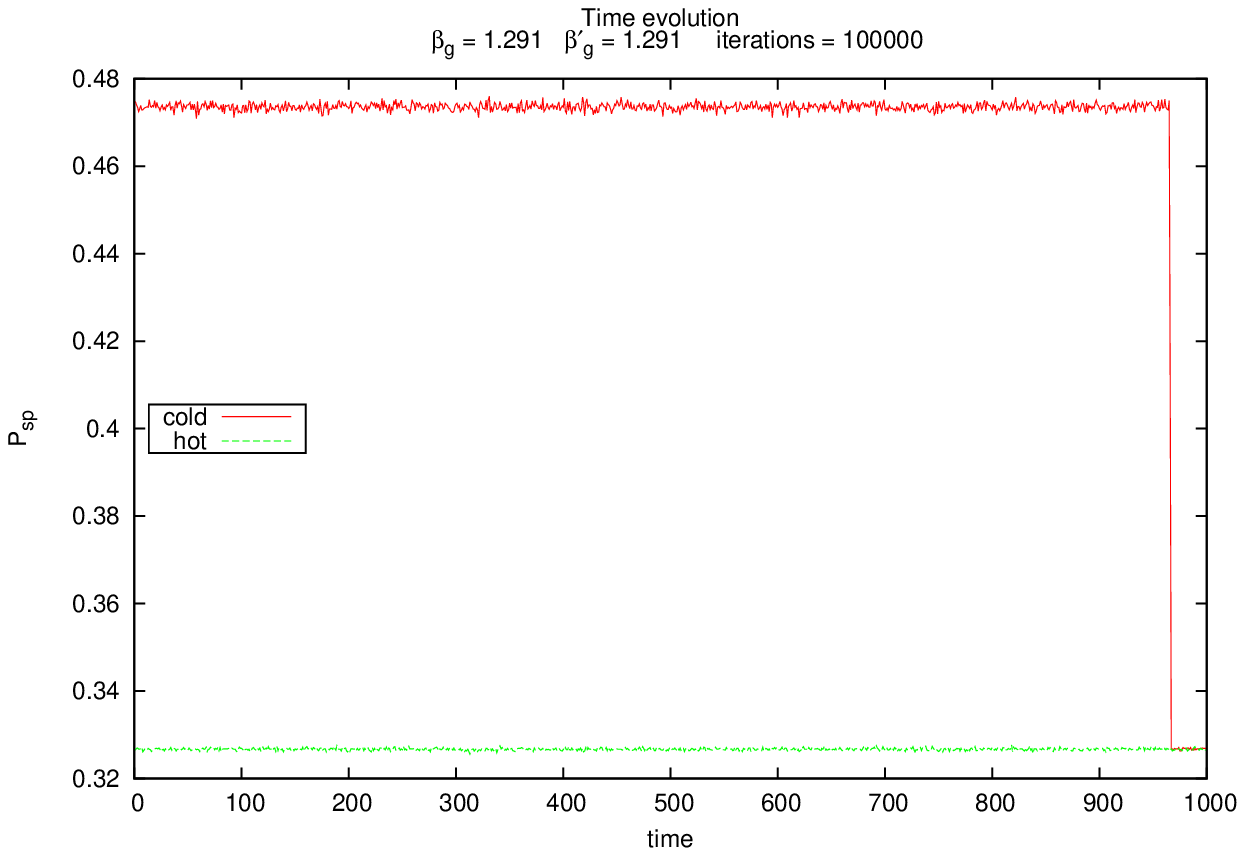,width=8.5cm}}
\end{minipage}
\caption{Long runs for six dimensions below the phase transition, at $\beta=1.289,\ 1.290$ and $1.291.$
\label{fig_6_129}}
\end{figure}
\begin{figure}[b]
\begin{minipage}{.49\textwidth}
\centerline{\epsfig{file=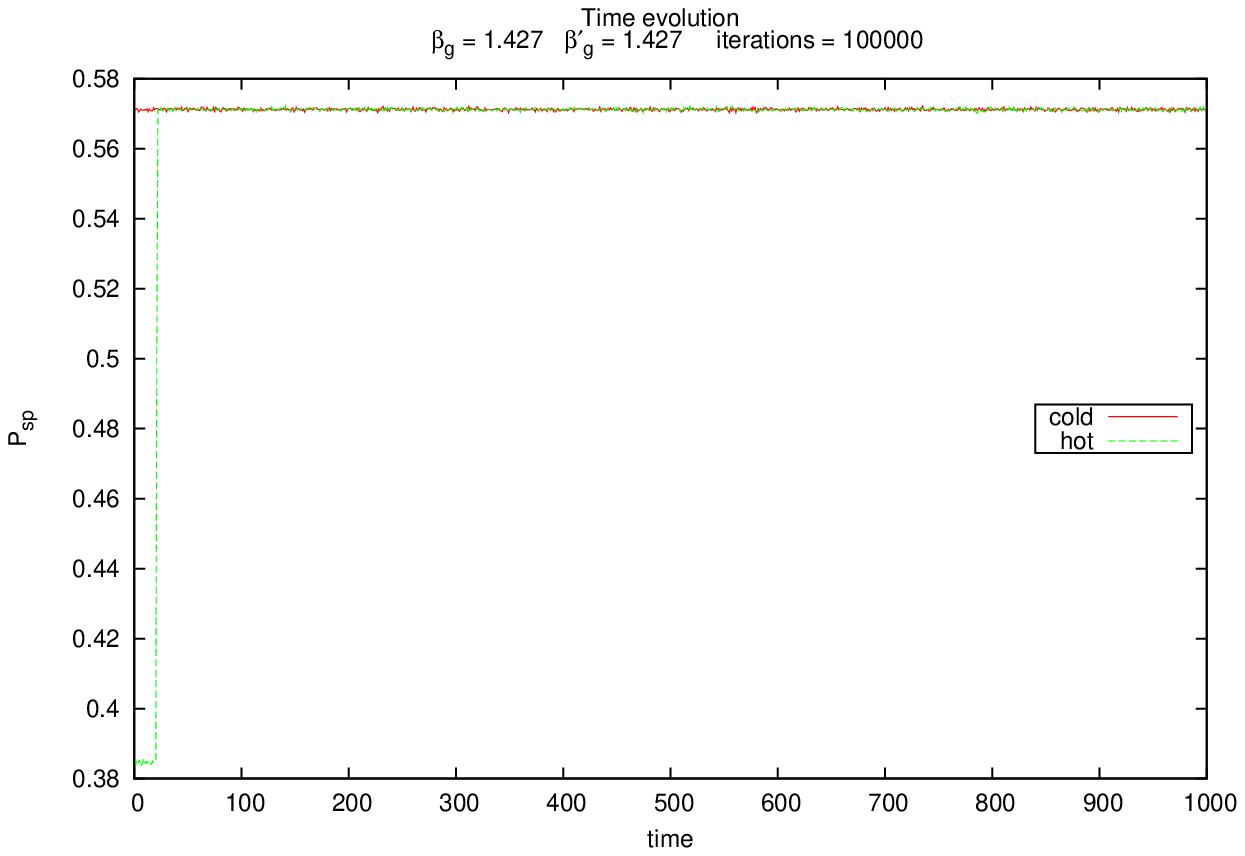,width=8.5cm}}
\end{minipage}
\begin{minipage}{.49\textwidth}
\centerline{\epsfig{file=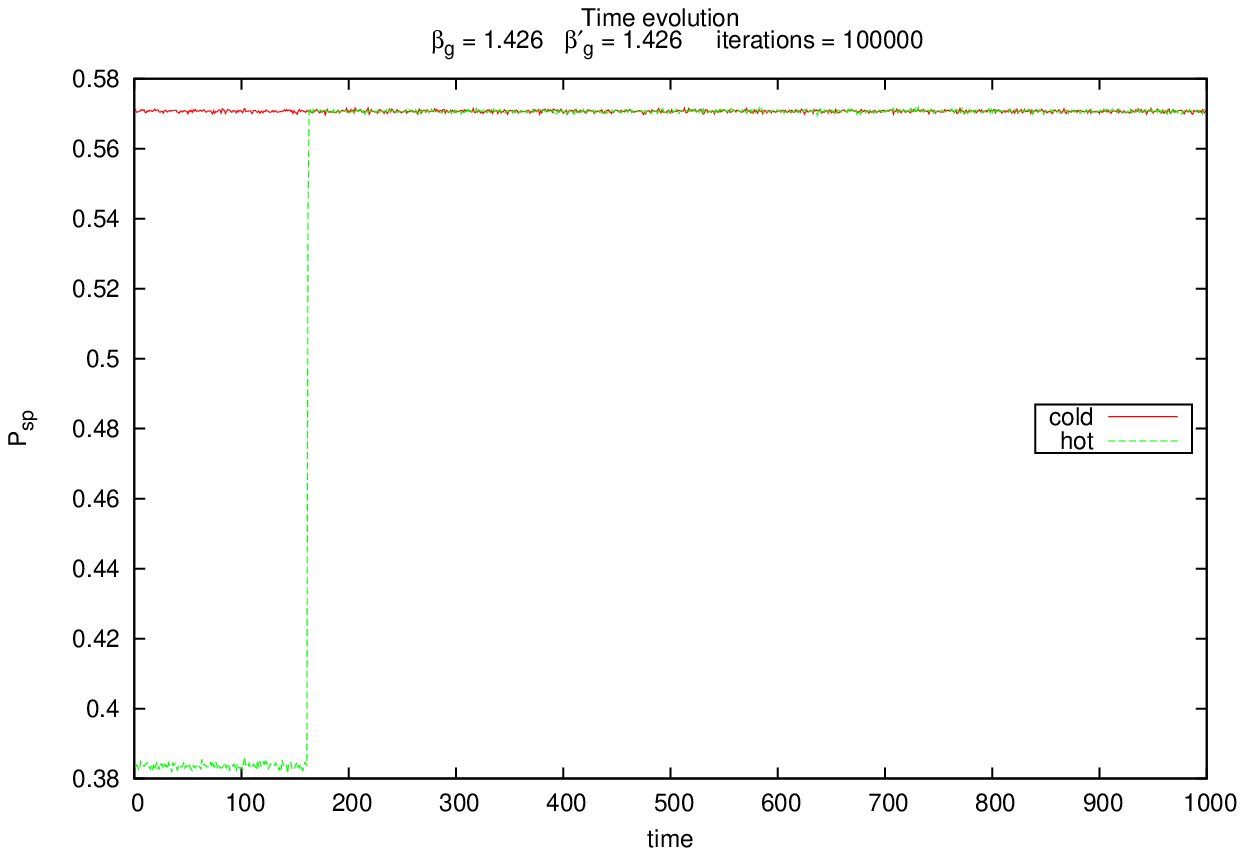,width=8.5cm}}
\end{minipage}
\begin{minipage}{.49\textwidth}
\centerline{\epsfig{file=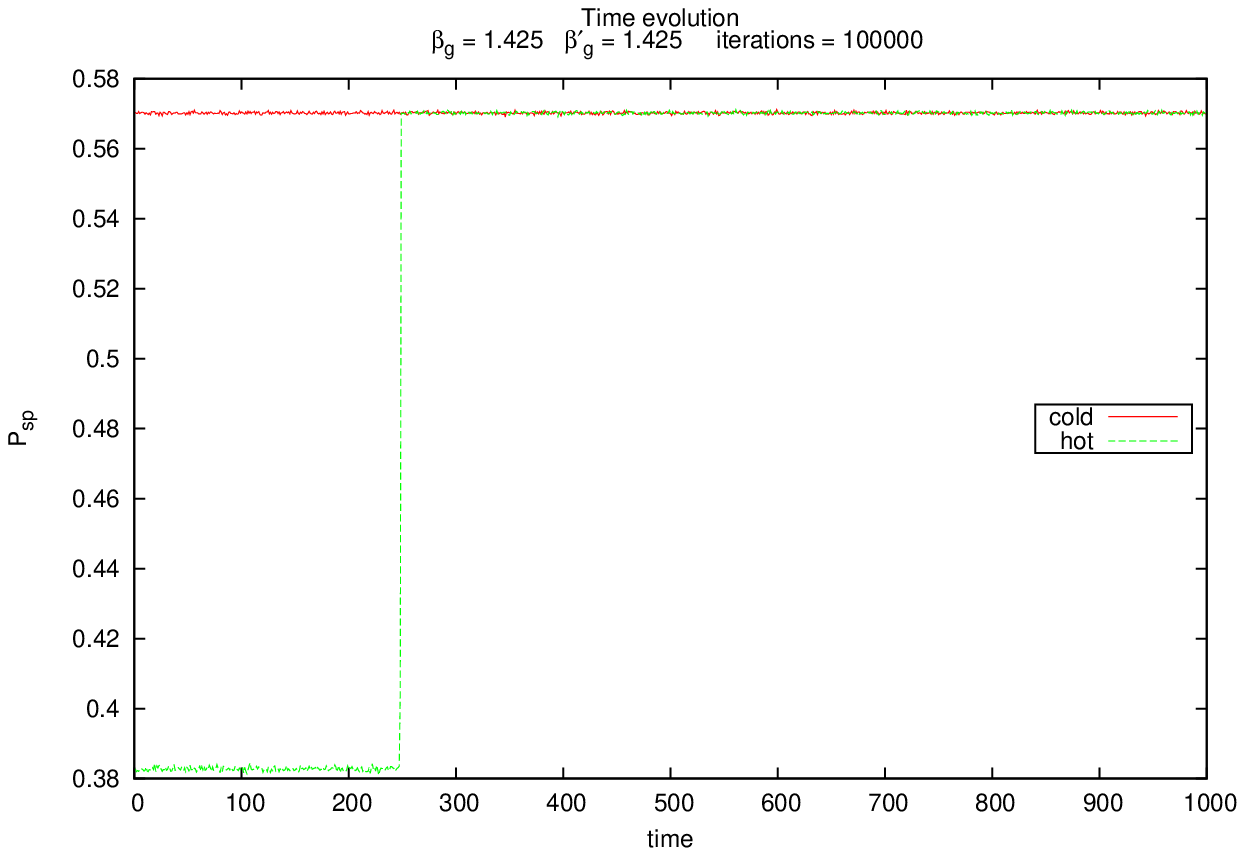,width=8.5cm}}
\end{minipage}
\caption{Long runs for six dimensions above the phase transition, at $\beta=1.427,\ 1.426$ and $1.425.$
\label{fig_6_142}}
\end{figure}
\end{landscape}

\begin{landscape}
\begin{figure}[t]
\begin{minipage}{.49\textwidth}
\centerline{\epsfig{file=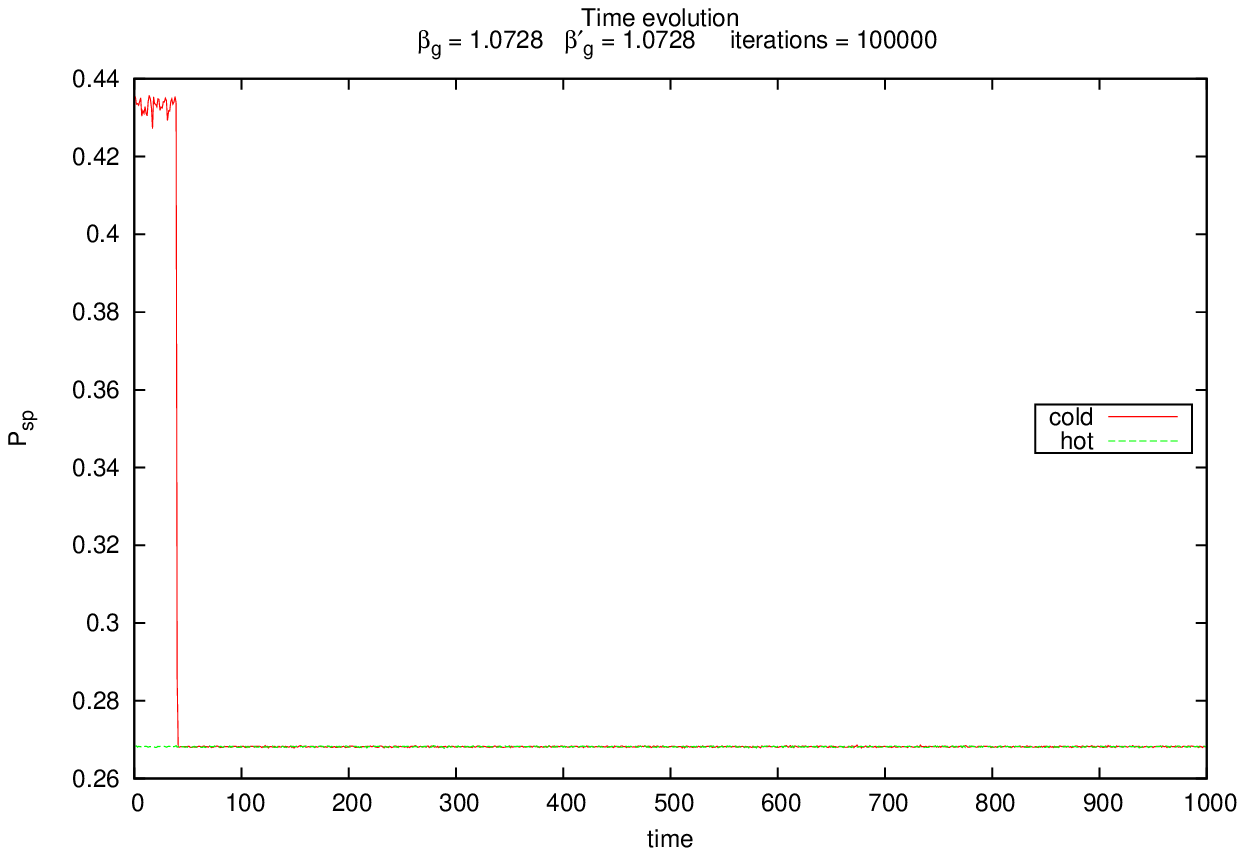,width=8.5cm}}
\end{minipage}
\begin{minipage}{.49\textwidth}
\centerline{\epsfig{file=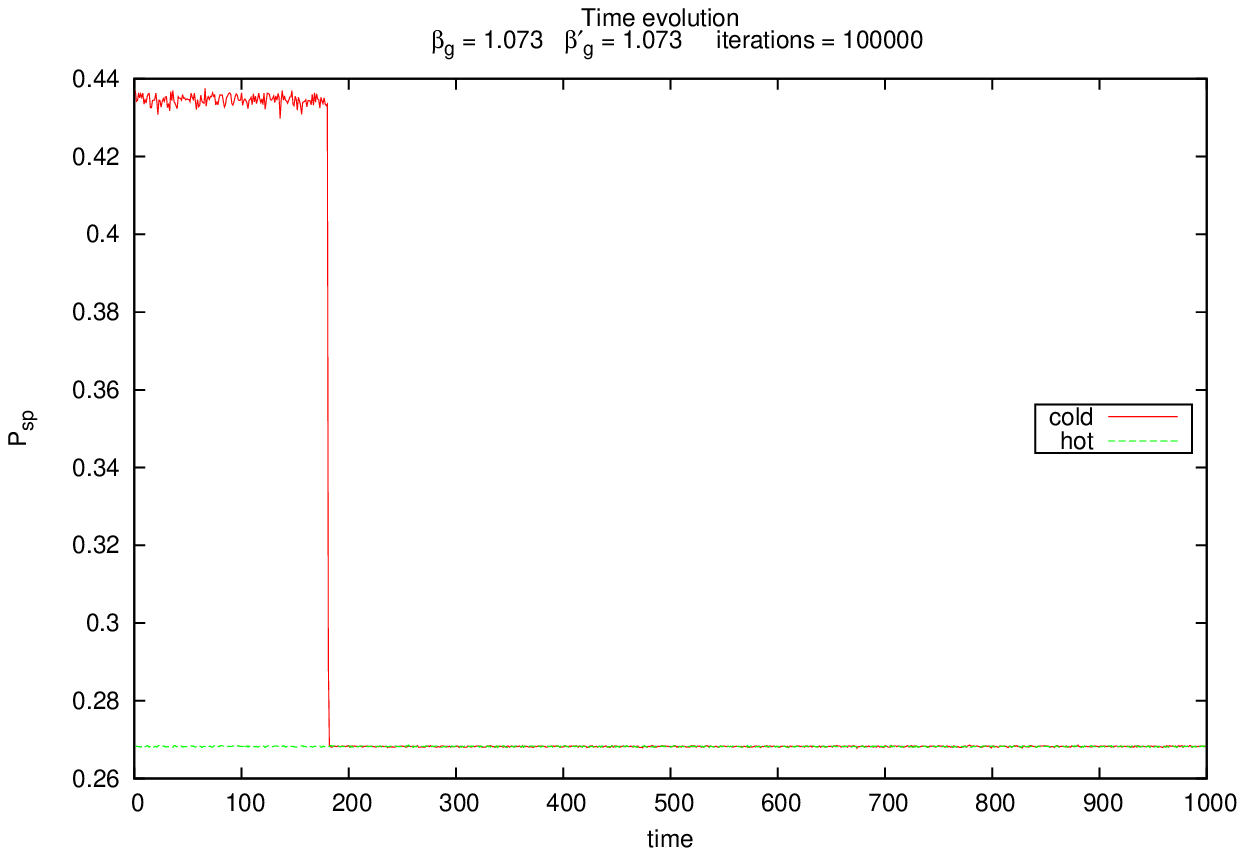,width=8.5cm}}
\end{minipage}
\begin{minipage}{.49\textwidth}
\centerline{\epsfig{file=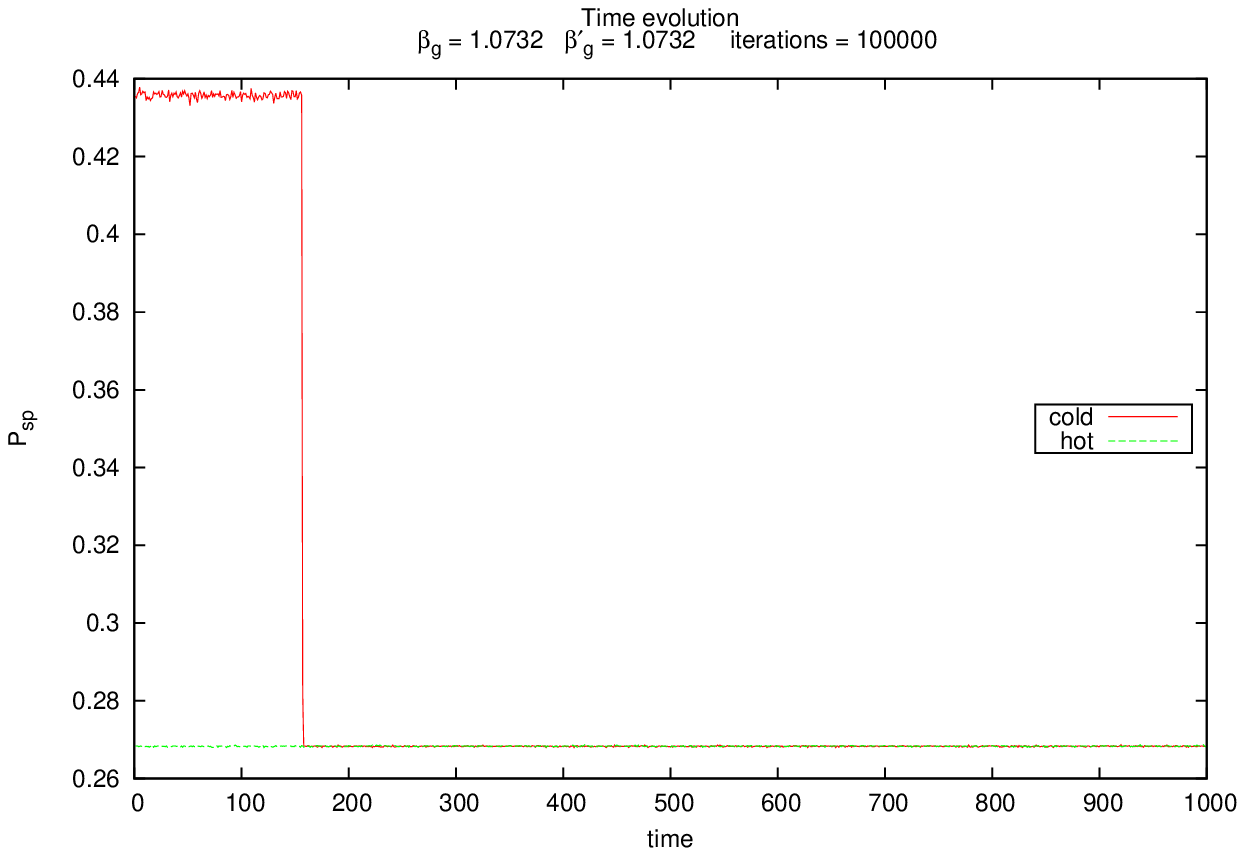,width=8.5cm}}
\end{minipage}
\caption{Long runs for seven dimensions below the phase transition, at $\beta=1.0728,\ 1.0730$ and $1.0732.$
\label{fig_7_107}}
\end{figure}
\begin{figure}[b]
\begin{minipage}{.49\textwidth}
\centerline{\epsfig{file=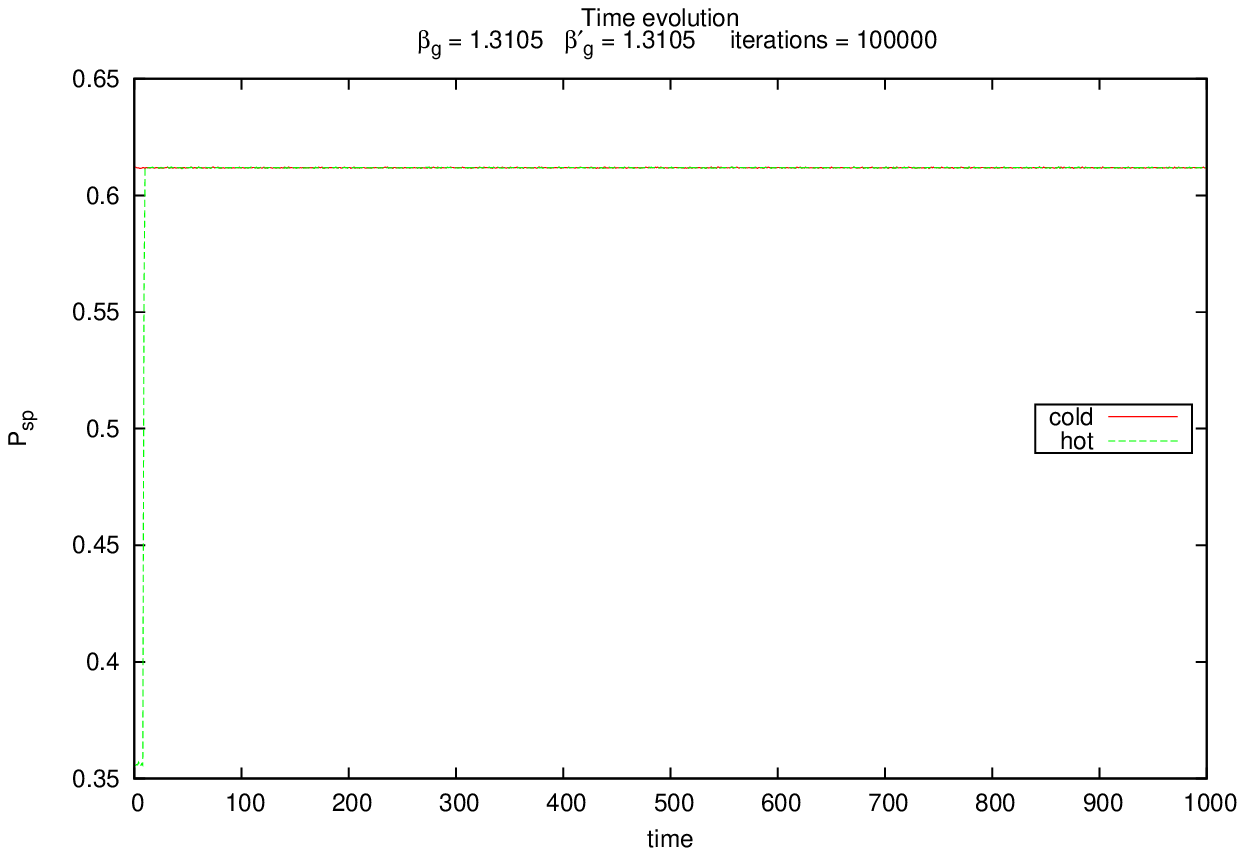,width=8.5cm}}
\end{minipage}
\begin{minipage}{.49\textwidth}
\centerline{\epsfig{file=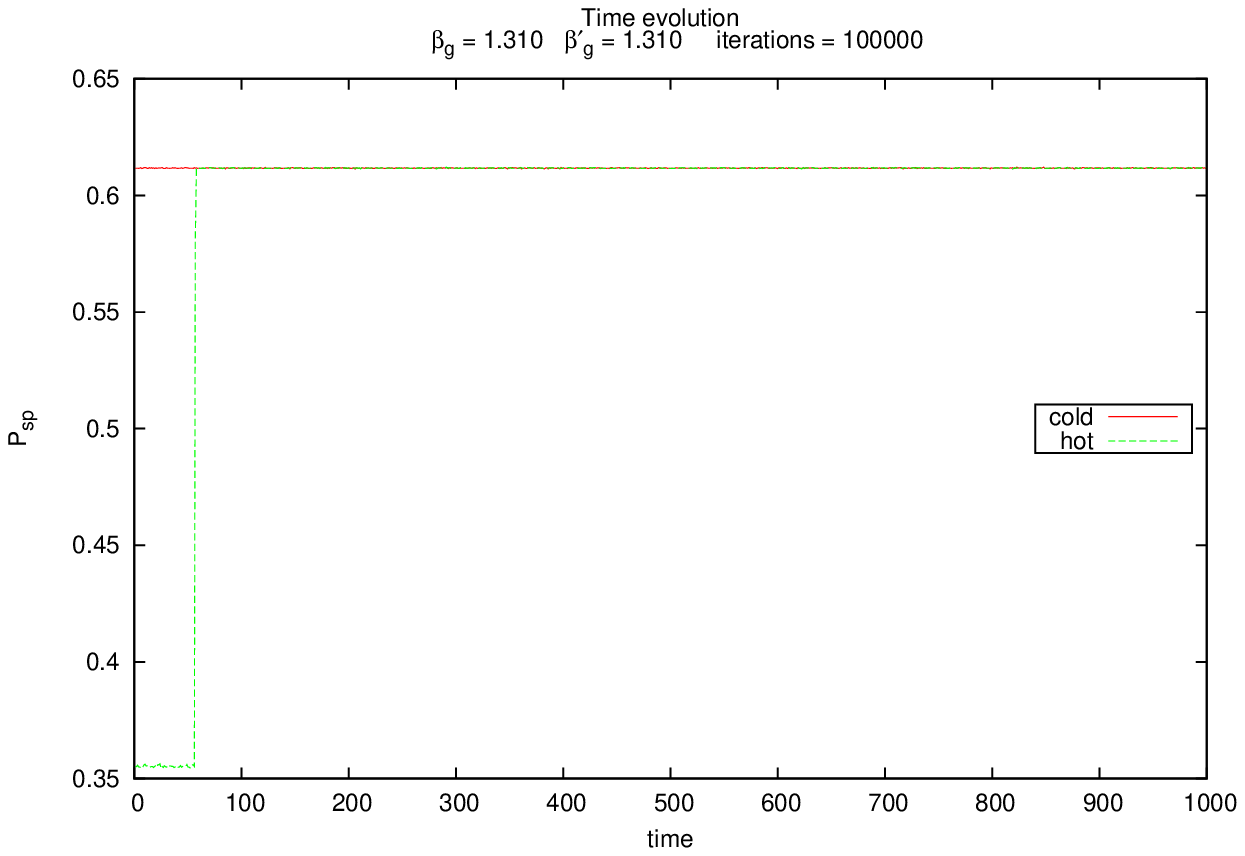,width=8.5cm}}
\end{minipage}
\begin{minipage}{.49\textwidth}
\centerline{\epsfig{file=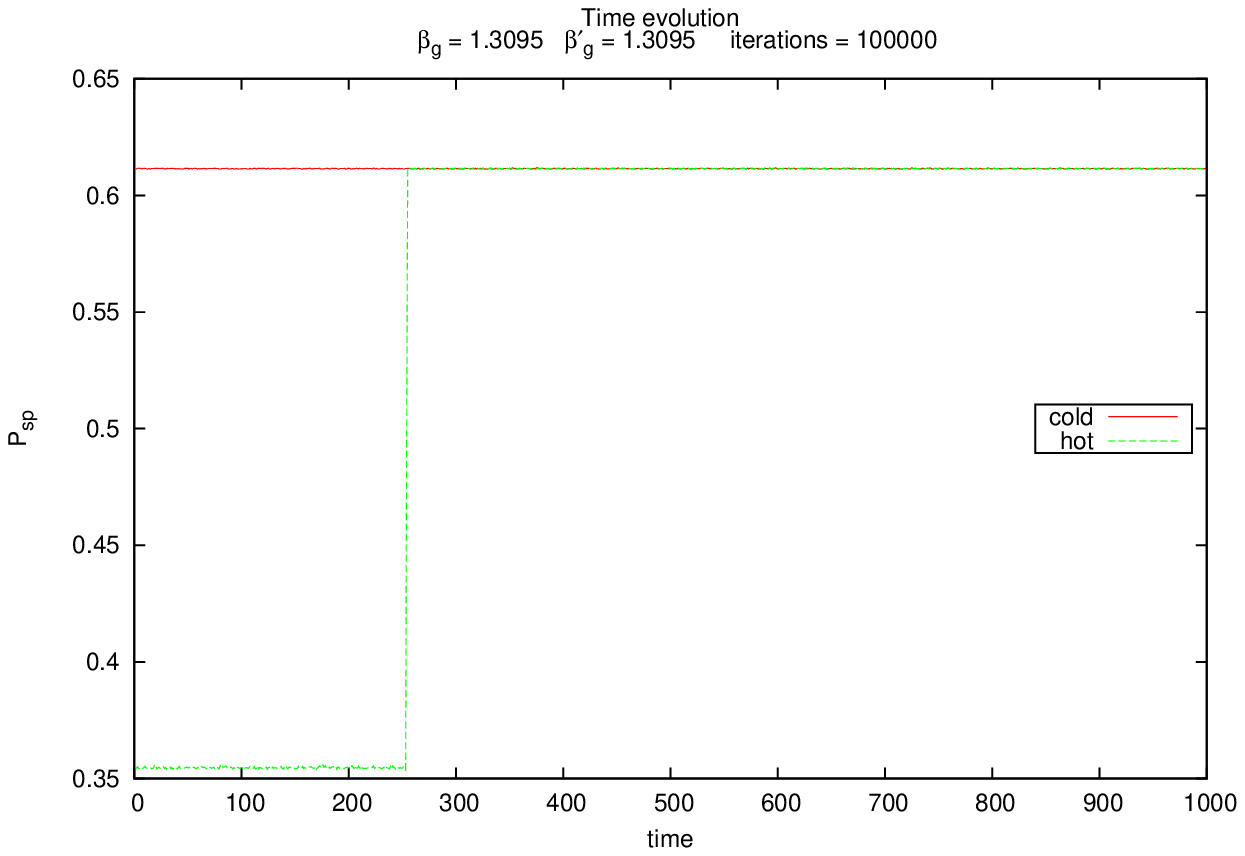,width=8.5cm}}
\end{minipage}
\caption{Long runs for seven dimensions above the phase transition, at $\beta=1.3105,\ 1.3100$ and $1.3095.$
\label{fig_7_131}}
\end{figure}
\end{landscape}

\begin{landscape}
\begin{figure}[h!]
\captionsetup{font=footnotesize}
\captionof{figure}{Plaquette values with errors in 6,7 and 8 dimensions }
\begin{center}
\includegraphics[scale=0.65]{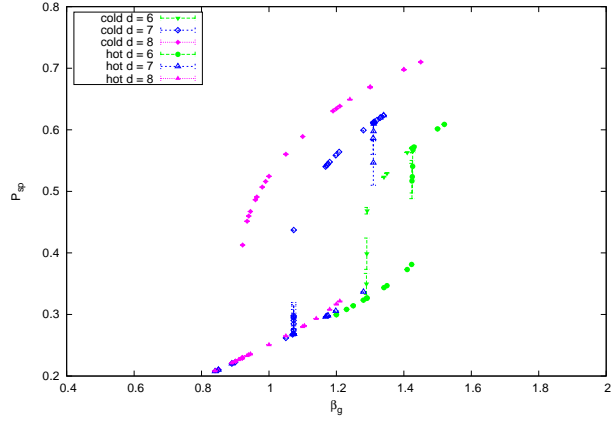}
\end{center}
\end{figure}
\begin{minipage}{0.5\textwidth}
\begin{tiny}
\centering
\begin{tabular}{|c|c|c|c|c|}
\hline 
$b_g$ & \tiny{$\substack{\text{Cold start}  \\ \text{values}}$} & \tiny{$\substack{\text{Cold start}  \\ \text{errors}}$} & \tiny{$\substack{\text{Hot start}  \\ \text{values}}$} & \tiny{$\substack{\text{Hot start}  \\ \text{errors}}$}  \\ \hline  1.200 & 0.299589 & 0.000011 & 0.299589 & 0.000011 \\ \hline 1.230 & 0.308219 & 0.000007 & 0.308219 & 0.000007 \\ \hline 1.250 & 0.314130 & 0.000007 & 0.314130 & 0.000007 \\ \hline 1.280 & 0.323297 & 0.000007 &0.323297 & 0.000007 \\ \hline 1.289 & 0.349 &0.017 & 0.32615 & 0.00001 \\ \hline 1.290 & 0.40 &0.03& 0.326446 & 0.000013 \\ \hline 1.291 &0.469 & 0.005 &0.326765  &  0.000004\\ \hline 1.341 & 0.523120 &0.000008  & 0.343620 & 0.000012 \\ \hline 1.350 &0.529227  &0.000012 &0.346879 &0.000013 \\ \hline 1.410 & 0.563112 & 0.000013 & 0.373005 & 0.000009 \\ \hline 1.423 & 0.569356 &0.000008 & 0.381212 & 0.000015 \\ \hline 1.424 &0.569824  &0.000011& 0.52& 0.03\\ \hline 1.425 & 0.570293 &  0.000009 & 0.52 & 0.03 \\ \hline 1.426 & 0.570760 & 0.000009 & 0.54 &0.02 \\ \hline 1.427 & 0.571259 & 0.000005 & 0.567 & 0.004 \\ \hline 1.430 &  0.572597 &  0.000008&0.5720& 0.0006 \\ \hline 1.500 & 0.60164 & 0.00014 & 0.601475  & 0.000009 \\ \hline 1.520 & 0.608791 & 0.000007 & 0.608775 & 0.000009 \\ 
\hline
\end{tabular}
\captionsetup{font=footnotesize}
\captionof{table}{Plaquette values \& errors  in 6 dimensions}
\end{tiny}
\end{minipage}
\begin{minipage}{0.5\textwidth}
\begin{center}
\begin{tiny}
\begin{tabular}{|c|c|c|c|c|}
\hline 
$b_g$ & \tiny{$\substack{\text{Cold start}  \\ \text{values}}$} & \tiny{$\substack{\text{Cold start}  \\ \text{errors}}$} & \tiny{$\substack{\text{Hot start}  \\ \text{values}}$} & \tiny{$\substack{\text{Hot start}  \\ \text{errors}}$}   \\ \hline 0.8500 &  0.210118 & 0.000003 & 0.210116 & 0.000005 \\ \hline 0.8900 & 0.220185 & 0.000005 & 0.220185 & 0.000003  \\ \hline 1.0700 &  0.267476  & 0.000003 & 0.267473 & 0.000005 \\ \hline
1.0728 & 0.275 & 0.007 & 0.268251 & 0.000003 \\ 
\hline 1.0730 & 0.30 & 0.02 & 0.268303 & 0.000006 \\ \hline 1.0731 &0.290 &0.018 &0.268336& 0.000003 \\ \hline 1.0732 & 0.30 & 0.02 & 0.268388 & 0.000009 \\ \hline 1.0733 & 0.285 & 0.017 &0.268393 &0.000006\\ \hline 1.0734 & 0.30& 0.02 & 0.268415& 0.000003 \\ \hline 1.0735 & 0.437191 & 0.000017 & 0.268445& 0.000003 \\ \hline 1.1680 & 0.540070 & 0.000003 & 0.296037 &0.000004 \\ \hline 1.1720 & 0.542651 & 0.000004 & 0.297304  & 0.000004 \\ \hline 1.1750 & 0.544582 & 0.000003 &0.298252 &0.000004  \\ \hline 1.1980 & 0.558431&  0.000005 & 0.305641 & 0.000005   \\ \hline 1.2800 & 0.599198 & 0.000005 &  0.336737&  0.000008  \\ \hline 1.3090& 0.611366& 0.000004& 0.59 & 0.03  \\ \hline 1.3095 &0.611544 &0.000002 &0.55 &0.04 \\ \hline 1.3100 &0.611763& 0.000002&  0.597& 0.015\\
\hline 1.3105 &0.611938& 0.000003&  0.609 &0.002\\
\hline 1.3120& 0.612546 & 0.000004 &  0.612549 & 0.000004\\
\hline 1.3300& 0.619573& 0.000003& 0.619576 &0.000004 \\
\hline 1.3400 &0.623345& 0.000004 & 0.623346 & 0.000004 \\
\hline
\end{tabular}
\captionsetup{font=footnotesize}
\captionof{table}{Plaquette values \& errors  in 7 dimensions}
\end{tiny}
\end{center}
\end{minipage}
\begin{minipage}{0.5\textwidth}
\begin{center}
\begin{tiny}
\begin{tabular}{|c|c|c|c|c|}
\hline 
$b_g$ & \tiny{$\substack{\text{Cold start}  \\ \text{values}}$} & \tiny{$\substack{\text{Cold start}  \\ \text{errors}}$} & \tiny{$\substack{\text{Hot start}  \\ \text{values}}$} & \tiny{$\substack{\text{Hot start}  \\ \text{errors}}$}   \\ \hline 0.8400& 0.208402 &0.000001 &0.208402 &0.000001\\ \hline 0.8900 &0.221268& 0.000001 & 0.221268& 0.000001 \\ \hline 0.9000 &0.223868& 0.000003 & 0.223873 &0.000003 \\ \hline
0.9100 &0.226487 &0.000001 &0.226487& 0.000001\\ 
\hline 0.9200& 0.229113 &0.000004& 0.229110 &0.000002\\ \hline 0.9215& 0.2301 &0.0006& 0.229511& 0.000004\\ \hline 0.9350 &0.451421& 0.000010 &0.233080 &0.000003\\ \hline 0.9400& 0.459860& 0.000006& 0.234412& 0.000002 \\ \hline 0.9450 &0.467337& 0.000011& 0.235743& 0.000004 \\ \hline 1.0000& 0.524477 &0.000005& 0.250721& 0.000003 \\ \hline 1.0500& 0.560293& 0.000004& 0.264967& 0.000003\\ \hline 1.1000 &0.589056& 0.000004  &0.280106 &0.000001 \\ \hline 1.2000 &0.634465 &0.000001  & 0.316682& 0.000002 \\ \hline 1.2100& 0.638365 &0.000006 & 0.321559& 0.000005 \\ \hline 1.3000 &0.669492& 0.000003  & 0.669492 &0.000003\\ \hline 1.4000& 0.697810 &0.000004 & 0.697810 &0.000004\\
\hline
\end{tabular}
\captionsetup{font=footnotesize}
\captionof{table}{Plaquette values \& errors  in 8 dimensions}
\end{tiny}
\end{center}
\end{minipage}
\end{landscape}


\end{document}